\documentclass[oribibl,envcountsame]{llncs}
 
\usepackage{latexsym}
\usepackage{amstext}
\usepackage{enumerate}

\title{Equivalence and Isomorphism
for Boolean Constraint Satisfaction\thanks{Supported
in part by grant 
NSF-INT-9815095/\protect\linebreak[0]DAAD-315-PPP-g\"u-ab.}}
\author{Elmar B\"ohler\inst{1}\thanks{Corresponding author.
Fax: +49-931-8886661}
\and Edith Hemaspaandra\inst{2}
\and Stef{}fen Reith\inst{3}
\and Heribert Vollmer\inst{1}}

\institute{Theoretische Informatik,
Universit\"{a}t W\"{u}rzburg,
Am Hubland,
D-97074 W\"{u}rzburg, Germany,
e-mail: [boehler,vollmer]@informatik.uni-wuerzburg.de
\and
Department of Computer Science,
Rochester Institute of Technology,
Rochester, NY 14623, U.S.A.,
e-mail: eh@cs.rit.edu
\footnote{Work done in part while visiting
Julius-Maximilians-Universit\"{a}t W\"{u}rzburg.}
\and
ActiveFilm,
Rosengasse 10,
D-97070 W\"{u}rzburg, Germany,
e-mail: reith@active-film.com
\footnote{Work done in part while employed at
Julius-Maximilians-Universit\"{a}t W\"{u}rzburg.}
}

\newtheorem{ourclaim}[theorem]{Claim}
\newcommand{\pair}[1]{\mbox{$(#1)$}}
\newcommand{\SAT}{\mbox{\rm CSP}}

\newcommand{\SATneqzero}{\SAT_{\neq \vec{0}}}
\newcommand{\SATneqone}{\SAT_{\neq \vec{1}}}
\newcommand{\SATneqzeroone}{\SAT_{\neq \vec{0}, \vec{1}}}
\newcommand{\EQUIV}{\mbox{\rm EQUIV}}
\newcommand{\ISO}{\mbox{\rm ISO}}

\newcommand{\GI}{\mbox{\rm GI}}
\newcommand{\VCGI}{\mbox{\rm VCGI}}
\newcommand{\condition}{\ |\ }
\newcommand{\littlep}{{p}}
\newcommand{\manyone}{\ensuremath{\,\leq_{m}^{{\littlep}}\,}}

\newcommand{\np}{{\rm NP}}
\newcommand{\p}{{\rm P}}
\newcommand{\parallelnp}{\mbox{$\p_{||}^{\np}$}}

\newcommand{\xor}{\oplus}

\newcommand{\set}[1]{\ensuremath{\{#1\}}} 
\newcommand{\enu}[3]{\ensuremath{{#1_{#2}}, \allowbreak \dots ,\allowbreak{#1_{#3}}}}
\newcommand{\cnto}[1]{\ensuremath{\#_{1}(#1)}}
\newcommand{\eqd}{\ensuremath{=_{\mathrm{\scriptscriptstyle def}}}}    

\begin{document}

\maketitle

\begin{abstract}
A Boolean constraint satisfaction instance is a conjunction of
constraint applications, where the allowed constraints are drawn from
a fixed set $\cal C$ of Boolean functions. We consider the problem of
determining whether two given constraint satisfaction instances are
equivalent and prove a \emph{Dichotomy Theorem} by showing that for
all sets $\cal C$ of allowed constraints, this problem is either
polynomial-time solvable or coNP-complete, and we give a simple
criterion to determine which case holds.

A more general problem addressed in this paper is the isomorphism
problem, the problem of determining whether there exists a renaming of
the variables that makes two given constraint satisfaction instances
equivalent in the above sense. We prove that this problem is coNP-hard
if the corresponding equivalence problem is coNP-hard, and
polynomial-time many-one reducible to the graph isomorphism problem in
all other cases.

\smallskip


\noindent\textbf{Track:} A
\end{abstract}


\sloppy

\section{Introduction}

In 1978, Thomas J.~Schaefer proved a remarkable result.  He examined
the satisfiability of propositional formulas for certain syntactically
restricted formula classes. Each such class is given by the finite
set $\cal C$
of Boolean functions allowed when constructing formulas.  A
\emph{$\cal C$-formula} in his sense now is a conjunction of clauses,
where each clause consists of a Boolean function from $\cal C$ applied
to some propositional variables. Such a Boolean function can be
interpreted as a constraint that has to be fulfilled by a given
assignment; the satisfiability problem for $\cal C$-formulas hence
provides a mathematical model for the examination of the complexity of
constraint satisfaction problems, studied in artificial intelligence
and database theory. Let $\SAT({\cal C})$ denote the problem of deciding
whether a given $\cal C$-formula is satisfiable.  Schaefer~\cite{Sch78}
showed that, depending on $\cal C$, the problem $\SAT({\cal C})$ is either
(1) efficiently (i.e., in polynomial time) solvable or (2)
NP-complete (see also \cite[Problem LO6]{GaJo79}); and he gave a
simple criterion that allows one to determine whether (1) or (2)
holds.  Since the complexity of $\SAT({\cal C})$ is either easy or
hard (and not located in one of the---under the assumption P $\neq$
NP---infinitely many intermediate degrees between P and the
NP-complete sets \cite{Lad75}), Schaefer called this a ``dichotomy
theorem for satisfiability.''

In recent years, there has been renewed interest in Schaefer's
result and constraint satisfaction problems.\\
N.~Creignou examined in \cite{Cr95} how difficult it is to find
assignments to constraint satisfaction problems that do not
necessarily satisfy all clauses but as much as possible.
Together with Hermann she studied the difficulty of determining the
number of satisfying assignments of a given constraint satisfaction
problem  in \cite{CrHe96}. 
In \cite{CrHe97} Creignou and H\'erbrard discussed algorithms that generate
all satisfying assignments, turning their attention to
the question whether such an algorithm, given the fact that it has
already found a satisfying assignment, can find another one in
polynomial time.
Kirousis and Kolaitis researched the complexity of finding minimal
satisfying assignments for constraint satisfaction problems in 
\cite{KiKo01} and
Khanna, Sudan and Trevisan examined the approximability of
these problems \cite{KhSuTr97,KhSuWi97}.
Reith and Vollmer had a look at lexicographical minimal or maximal 
satisfying assignments constraint satisfaction
problems \cite{ReVo00}.
In \cite{ReWa99} Reith and Wagner examined very closely various problems in the
vicinity of constraint satisfaction such as the circuit value problem,
counting and threshold problems for restricted classes of Boolean circuits.
The PhD thesis of S.~Reith \cite{Re01} contains a wealth of results
about problems dealing with restricted Boolean circuits, formulas,
and constraint satisfaction.

As mentioned above, constraint satisfaction problems are used as a
programming or query language in fields such as artificial
intelligence and database theory, and the above complexity results
shed light on the difficulty of design of systems in that areas. A
problem of immense importance from a practical perspective is that
of determining whether two sets of constraints express the same state of
affairs (that is, are equivalent), for example, in the applications, if two
programs or queries are equivalent, or if a program matches a given
specification. Surprisingly, this problem has not yet been looked at
from a complexity point of view. In the case of unrestricted
propositional formulas, the equivalence problem is easily seen to be
complete for coNP.  The main result of the present paper
(Theorem~\ref{dichth}) is a complete classification of the complexity
of determining if two given constraint satisfaction instances are
equivalent.  We consider constraints drawn from a fixed arbitrary
finite set $\cal C$ of Boolean functions and show that for all such
$\cal C$, the considered problem is either (1) solvable in polynomial
time, or (2) complete for coNP.  As in Schaefer's result, our proof is
constructive in the sense that it allows us to easily determine, given
$\cal C$, if (1) or (2) holds.

Besides the immediate practical relevance of the equivalence problem,
we also see our results as contributions to the study of two other
decision problems: First, the equivalence problem is a ``sub-problem''
of the minimization problem, i.e., the problem to find out, given a
set of constraints, if it can equivalently be expressed with a fewer
number of constraints.  Secondly, equivalence relates to the isomorphism
problem, which has been studied from a theoretical perspective for various
mathematical structures. Most prominently, the question if two given
(directed or undirected) graphs are isomorphic is one of the few
problems in NP neither known to be in P nor known to be NP-complete
\cite{KoScTo93}.  The most recent news about graph isomorphism are a
number of hardness results (e.g., for NL, PL, and DET) given in
\cite{To00}.  Related to our study are the papers \cite{AgTh00,BRS95}
presenting a number of results concerning isomorphism of propositional
formulas. In Section~\ref{s:iso}, we show (Theorem~\ref{isodichth})
that the isomorphism problem for constraint applications is coNP-hard
if the corresponding equivalence problem is coNP-hard, and
polynomial-time many-one reducible to the just mentioned graph
isomorphism problem in all other cases. 
We also show that for a number of these cases,
the isomorphism problem is in fact
polynomial-time many-one equivalent to graph isomorphism
(Theorems~\ref{giequivth} and \ref{t:gi-affine}).
The same proof technique can be used to prove a general,
non-trivial $\parallelnp$ (parallel access to NP) upper bound 
for the isomorphism problems for constraint satisfaction
(Theorem~\ref{t:iso-upper}).

\section{Boolean Constraint Satisfaction Problems}

We start by formally introducing constraint satisfaction problems.
The definitions necessary for the equivalence and isomorphism problems
will be given in the upcoming sections.

\begin{definition}\rm
\begin{enumerate}
\item A {\em constraint} $C$ (of arity $k$) is a Boolean function from 
$\{0,1\}^k$ to $\{0,1\}$.
\item If $C$ is a constraint of arity $k$, and $x_1, x_2, \dots, x_k$
are (not necessarily distinct) variables, then $C(x_1, x_2, \dots,
x_k)$ is a {\em constraint application of $C$}.
\item If $C$ is a constraint of arity $k$, and for $1 \leq i \leq k$,
$x_i$ is a variable or a constant (0 or 1), then $C(x_1, x_2, \dots,
x_k)$ is a constraint application of $C$ {\em with constants}.
\end{enumerate}
\end{definition}

The decision problems examined by Schaefer are the following.

\begin{definition}\rm
Let ${\cal C}$ be a finite set of constraints.
\begin{enumerate}
\item $\SAT({\cal C})$ is the problem of, given a set $S$ of
constraint applications of ${\cal C}$, to decide whether $S$ is
satisfiable, i.e., whether there exists an assignment to the variables
of $S$ that satisfies every constraint application in $S$.
\item $\SAT_c({\cal C})$ is the problem of, given a set $S$ of
constraint applications of ${\cal C}$ with constants, to decide
whether $S$ is satisfiable.
\end{enumerate}
\end{definition}

Clearly, there are an infinite number of $\SAT({\cal C})$ problems.
In 1978, Schaefer proved the surprising result that constraint
satisfiability problems are either in P or NP-complete.  He also
completely characterized for which sets of constraints the problem is
in P and for which it is NP-complete.  Consult the excellent monograph
\cite{CrKhSu00} for an almost completely up-to-date overview of
further results and dichotomy theorems for constraint satisfaction
problems.

The question of whether satisfiability for CSPs is in P or NP-complete
depends on those properties of the involved Boolean functions that we
define next.

\begin{definition}\rm
Let $C$ be a constraint.
\begin{itemize}
\item $C$ is {\em 0-valid} if $C(\vec{0}) = 1$.
\item $C$ is {\em 1-valid} if $C(\vec{1}) = 1$.
\item $C$ is {\em Horn} (a.k.a.\ weakly negative) if $C$ is 
equivalent to a CNF formula where each clause has
at most one positive variable.
\item $C$ is {\em anti-Horn} (a.k.a.\ weakly positive) if $C$ is 
equivalent to a CNF formula where each clause has
at most one negative variable.
\item $C$ is {\em bijunctive} if $C$ is equivalent to a 2CNF formula.
\item $C$ is {\em affine} if $C$ is equivalent to an XOR-CNF formula.
\item $C$ is {\em complementive} (a.k.a.\ C-closed) if for every $s
\in \{0,1\}^k$, $C(s) = C(\overline{s})$, where $k$ is the arity of
$C$ and $\overline{s} \eqd (1-s_1)(1-s_2) \cdots (1-s_k)$ for $s =
s_1s_2 \cdots s_k$.
\end{itemize}
Let $\cal C$ be a finite set of constraints.  We say ${\cal C}$ is
0-valid, 1-valid, Horn, anti-Horn, bijunctive, affine, or
complementive if {\em every} constraint $C\in{\cal C}$ is 0-valid,
1-valid, Horn, anti-Horn, bijunctive, affine, or complementive,
respectively. Finally, we say that ${\cal C}$ is {\em Schaefer} if
${\cal C}$ is Horn or anti-Horn or affine or bijunctive.
\end{definition}

Schaefer's theorem can now be stated as follows.

\begin{theorem}[Schaefer \cite{Sch78}]
Let ${\cal C}$ be a finite set of constraints.
\begin{enumerate}
\item If ${\cal C}$ is 0-valid, 1-valid, or Schaefer, then $\SAT({\cal
C})$ is in {\rm P}; otherwise, $\SAT({\cal C})$ is {\rm NP}-complete.
\item If ${\cal C}$ is Schaefer, then $\SAT_c({\cal C})$ is in {\rm
P}; otherwise, $\SAT_c({\cal C})$ is {\rm NP}-complete.
\end{enumerate}
\end{theorem}

In this paper, we will study two other decision problems for
constraint satisfaction problems. In the next section, we will look at
the question of whether two given CSPs are equivalent.  In
Section~\ref{s:iso}, we address the isomorphism problem for CSPs.  In
both cases, we will prove dichotomy theorems.

\section{The Equivalence Problem for Constraint Satisfaction}

The decision problems studied in this section are the following:

\begin{definition}\rm
Let ${\cal C}$ be a finite set of constraints.
\begin{enumerate}
\item $\EQUIV({\cal C})$ is the problem of, given two sets $S$ and $U$
of constraint applications of ${\cal C}$, to decide whether $S$ and
$U$ are equivalent, i.e., whether for every assignment to the
variables, $S$ is satisfied if and only if $U$ is satisfied.
\item $\EQUIV_c({\cal C})$ is the problem of, given two sets $S$ and
$U$ of constraint applications of ${\cal C}$ with constants, to decide
whether $S$ and $U$ are equivalent.
\end{enumerate}
\end{definition}

It is immediate that all equivalence problems are in coNP.
Note that, in some sense, equivalence is at least as hard as
non-satisfiability, since $S$ is not satisfiable if and only if $S$
is equivalent to $0$. Thus, we obtain immediately that
if ${\cal C}$ is not Schaefer, then $\EQUIV_c({\cal C})$ is coNP-complete.

On the other hand, equivalence can be harder than satisfiability.
For example, equivalence between
Boolean formulas with $\wedge$ and $\vee$ (i.e., without negation) is
coNP-complete~\cite{EG95} while non-satisfiability for
these formulas is clearly in P.
We will prove following dichotomy theorem.

\begin{theorem}\label{dichth}
Let ${\cal C}$ be a set of constraints.  If ${\cal C}$ is Schaefer,
then $\EQUIV({\cal C})$ and $\EQUIV_c({\cal C})$ are in {\rm P};
otherwise, $\EQUIV({\cal C})$ and $\EQUIV_c({\cal C})$ are {\rm
coNP}-complete.
\end{theorem}

The cases of constraints with polynomial-time equivalence problems are easy
to identify, using the following theorem:

\begin{theorem}
\label{equivttsatth}
$\EQUIV_c({\cal C})$ is truth-table reducible to
$\SAT_c({\cal C})$.
\end{theorem}

\proof
Let $S$ and $U$ be two sets of constraint applications of ${\cal C}$
with constants.  Note that $S$ and $U$ are equivalent if and only if
$U \rightarrow A$ for every constraint application $A \in S$, and $S
\rightarrow B$ for every constraint application $B \in U$ (see, e.g.,
\cite{hako92}). Here and in the following, when we write a set of
constraint applications $S$ in a Boolean formula, we take this to
be a shorthand for $\bigwedge_{\widehat{A} \in S} \widehat{A}$.

Given a constraint application $\widehat{A}$ with constants and a set
$\widehat{S}$ of constraint applications of ${\cal C}$ with constants,  
it is easy to check whether $\widehat{S} \rightarrow \widehat{A}$ with
at most $2^k$ truth-table queries to $\SAT_c({\cal C})$,
where $k$ is the maximum arity of ${\cal C}$: 
For every assignment to the variables in $\widehat{A}$
that does not satisfy $\widehat{A}$, substitute this partial truth
assignment in $\widehat{S}$. $\widehat{S} \rightarrow \widehat{A}$ if and
only if none of these substitutions results in a satisfiable set
of constraint applications. 
\qed

If ${\cal C}$ is Schaefer, then $\SAT_c{(\cal C})$ is in P by Schaefer's
theorem and we immediately obtain the following corollary.

\begin{corollary}
If ${\cal C}$ is Schaefer, then $\EQUIV_c({\cal C})$ is in {\rm P}.
\end{corollary}

\label{s:lowerbound}

Having identified the easy equivalence cases, the following theorem
proves the second half of our Dichotomy Theorem~\ref{dichth}:

\begin{theorem}\label{eqhard}
If ${\cal C}$ is not Schaefer, then $\EQUIV({\cal C})$ is {\rm
coNP}-hard.
\end{theorem}

First of all, note that this would be easy to prove if
we had constants in the language, since for all sets $S$
of  constraint applications of ${\cal C}$ the following holds:
$S$ is not satisfiable
if and only if $S$ is equivalent to the constraint $0$. Still,
we can use this simple observation in the case where
${\cal C}$ is not 0-valid and not 1-valid.

\begin{ourclaim}
\label{NotZOVal}
If ${\cal C}$ is not Schaefer, not 0-valid, and not 1-valid, then
$\EQUIV({\cal C})$ is {\rm coNP}-hard.
\end{ourclaim}

\proof
We will reduce $\overline{\SAT({\cal C})}$ to $\EQUIV({\cal C})$.  Let
$S$ be a set of constraint applications of ${\cal C}$.  As noted
above, $S$ is not satisfiable if and only if $S$ is equivalent to
$0$. Let $C_0 \in {\cal C}$ be a constraint that is not 0-valid, and
let $C_1 \in {\cal C}$ be a constraint that is not 1-valid.  Note
that, for any variable $x$, $\{C_0(x, \dots, x), C_1(x, \dots, x)\}$
is equivalent to 0, and thus $S \in \overline{\SAT({\cal C})}$ if and
only $S$ is equivalent to $\{C_0(y, \dots, y), C_1(y, \dots, y)\}$.
\qed

If ${\cal C}$ is not Schaefer, but is 0-valid or 1-valid, then every
set of constraint applications of ${\cal C}$ is trivially satisfiable
(by $\vec{0}$ or $\vec{1}$). In these cases, a reduction from
$\overline{\SAT({\cal C})}$ will not help, since $\overline{\SAT({\cal
C})}$ is in P. However, we will show that in these cases the problem
of determining whether there exists a non-trivial satisfying
assignment is NP-complete and we will use the complements of these
satisfiability problems to reduce from.

Creignou and H\'ebrard prove the following result, concerning the
existence of non-trivial satisfying assignments (\cite[Proposition
4.7]{CrHe97}, their notation for our $\SATneqzeroone$ is ${\rm SAT}^*$):

\begin{proposition}[\cite{CrHe97}]
\label{t:SATneqzeroone}
If ${\cal C}$ is not Schaefer, then $\SATneqzeroone({\cal C})$ is {\rm
NP}-complete, where $\SATneqzeroone({\cal C})$ is the problem of,
given a set $S$ of constraint applications of ${\cal C}$, to decide
whether there is a satisfying assignment for $S$ other than $\vec{0}$
and $\vec{1}$.
\end{proposition}

$\SATneqzeroone({\cal C})$ corresponds to the notion of ``having
a non-trivial satisfying
assignment'' in the case that ${\cal C}$ is 0-valid and 1-valid. We
will reduce $\overline{\SATneqzeroone({\cal C})}$ to $\EQUIV({\cal
C})$ in this case in the proof of Claim~\ref{cl:zeroonevalid} to
follow.

For the cases that ${\cal C}$ is not 1-valid or not 0-valid, we obtain
the following analogues of Proposition~\ref{t:SATneqzeroone}.
A proof can be found in the appendix.

\begin{theorem}
\label{t:SATneqzero}
\begin{enumerate}
\item
If ${\cal C}$ is not Schaefer and not 0-valid then $\SATneqone({\cal
C})$ is {\rm NP}-complete, where $\SATneqone({\cal C})$ is the problem
of, given a set $S$ of constraint applications of ${\cal C}$, to
decide whether there is a satisfying assignment for $S$ other than
$\vec{1}$.
\item
If ${\cal C}$ is not Schaefer and not 1-valid then $\SATneqzero({\cal
C})$ is {\rm NP}-complete, where $\SATneqzero({\cal C})$ is the
problem of, given a set $S$ of constraint applications of ${\cal C}$,
to decide whether there is a satisfying assignment for $S$ other than
$\vec{0}$.
\end{enumerate}
\end{theorem}

\proof
Careful inspection of Creignou and H\'ebrard's proof of
Proposition~\ref{t:SATneqzeroone} shows that following holds if ${\cal C}$
is not Schaefer:
\begin{enumerate}
\item If ${\cal C}$ is not 0-valid and not 1-valid, then ${\cal L} =
\{S \condition S \in \SATneqzeroone({\cal C})$
and not $S(\vec{0})$ and not $S(\vec{1})\}$ is NP-complete
(this is case 1 of Creignou and H\'ebrard's proof).
\item If ${\cal C}$ is 0-valid and not 1-valid, then ${\cal L}_0 =
\{S \condition S \in \SATneqzeroone({\cal C})$
and not $S(\vec{1})\}$ is NP-complete (this is case 2b of
Creignou and H\'ebrard's proof).
\item If ${\cal C}$ is 1-valid and not 0-valid, then ${\cal L}_1 =
\{S \condition S \in \SATneqzeroone({\cal C})$
and not $S(\vec{0})\}$ is NP-complete (this is case 3b of
Creignou and H\'ebrard's proof).
\end{enumerate}

This almost immediately implies Theorem~\ref{t:SATneqzero}.  Let
${\cal C}$ be not Schaefer and not 0-valid.  If ${\cal C}$ is not
1-valid, then ${\cal L}$ trivially many-one reduces to
$\SATneqone({\cal C})$, since, for $S$ a set of constraint
applications of ${\cal C}$, $S \in {\cal L}$ if and only if not
$S(\vec{0})$, not $S(\vec{1})$,
and $S \in \SATneqone({\cal C})$.  Similarly, if ${\cal
C}$ is 1-valid, then ${\cal L}_1$ trivially many-one reduces to
$\SATneqone({\cal C})$.  This proves part (1) of
Theorem~\ref{t:SATneqzero}. Part (2) follows by symmetry.
\qed

\begin{ourclaim}\label{NotZoOVal}
Let ${\cal C}$ be a finite set of constraints.
\begin{enumerate}
\item If ${\cal C}$ is 1-valid, not Schaefer, and not 0-valid, then
$\EQUIV({\cal C})$ is {\rm coNP}-hard.
\item If ${\cal C}$ is 0-valid, not Schaefer and not 1-valid, then
$\EQUIV({\cal C})$ is {\rm coNP}-hard.
\end{enumerate}
\end{ourclaim}

\proof
We will prove the first case; the proof of the second case is similar.
We will reduce $\overline{\SATneqone({\cal C})}$ to $\EQUIV({\cal C})$
as follows.  Let $S$ be a set of constraint applications of ${\cal C}$
and let $x_1, \dots ,x_n$ be the variables occurring in $S$.  Note
that $\vec{1}$ satisfies $S$, since every constraint in $S$ is
1-valid. Therefore, $S \not \in \SATneqone({\cal C})$ if and only if
$S$ is equivalent to $\bigwedge_{i = 1}^n x_i$.  Let $C \in {\cal C}$
be not 0-valid. Since $C$ is 1-valid, $x_i$ is equivalent to $C(x_i,
\dots, x_i)$.  It follows that $S \not \in \SATneqone({\cal C})$ if
and only if $S$ is equivalent to $\{C(x_i, \dots, x_i) \condition 1
\leq i \leq n\}$.
\qed

The final case is where ${\cal C}$ is both 0-valid and 1-valid.  We
need the following key lemma from Creignou and H\'ebrard which is used
in their proof of Proposition~\ref{t:SATneqzeroone}.

\begin{lemma}[\cite{CrHe97}, Lemma 4.9(1)]
\label{l:CHkeylemma}
Let ${\cal C}$ be a set of constraints that is not Horn, not
anti-Horn, not affine, and 0-valid. Then either
\begin{enumerate}
\item There exists a set $V_0$ of constraint applications of ${\cal
C}$ with variables $x$ and $y$ and constant $0$ such that $V_0$ is
equivalent to $x \rightarrow y$, or
\item There exists a set $V_0$ of constraint applications of ${\cal
C}$ with variables $x, y, z$ and constant $0$ such that $V_0$ is
equivalent to $(\overline{x} \wedge \overline{y} \wedge \overline{z})
\vee (x \wedge \overline{y} \wedge z) \vee (\overline{x} \wedge y
\wedge z)$.
\end{enumerate}
\end{lemma}

\begin{ourclaim}
\label{cl:zeroonevalid}
Let ${\cal C}$ be a finite set of constraints.  If ${\cal C}$ is not
Schaefer but both 0-valid and 1-valid, then $\EQUIV({\cal C})$ is
{\rm coNP}-hard.
\end{ourclaim}

\proof
We will reduce $\overline{\SATneqzeroone({\cal C})}$ to $\EQUIV({\cal
C})$. Let $S$ be a set of constraint applications of ${\cal C}$ and
let $x_1, \dots x_n$ be the variables occurring in $S$.  Note that
$\vec{0}$ and $\vec{1}$ satisfy $S$, since every constraint in $S$ is
0-valid and 1-valid. Therefore, $S \not \in \SATneqzeroone({\cal C})$
if and only if $S$ is equivalent to $\bigwedge_{i = 1}^n x_i \vee
\bigwedge_{i = 1}^n \overline{x_i}$.

First, suppose there is a constraint $C \in {\cal C}$ that is
non-complementive.  (This case is similar to Creignou and H\'ebrard's
case 2a).  Let $k$ be the arity of $C$ and let $s \in \{0,1\}^k$ be an
assignment such that $C(s) = 1$ and $C(\overline{s}) = 0$.  Let
$A(x,y)$ be the constraint application $C(a_1, \dots, a_k)$, where
$a_i = y$ if $s_i = 1$ and $a_i = x$ if $s_i = 0$.  Then $A(0,0) =
A(1, 1) = 1$, since $A$ is 0-valid and 1-valid; $A(0, 1) = 1$, since
$C(s) = 1$; and $A(1, 0) = 0$, since $C(\overline{s}) = 0$.  Thus,
$A(x,y)$ is equivalent to $x \rightarrow y$.  Since $\bigwedge_{i =
1}^n x_i \vee \bigwedge_{i = 1}^n \overline{x_i}$ is equivalent to
$\bigwedge_{1 \leq i,j \leq n} (x_i \rightarrow x_j)$, it follows that
$S \not \in \SATneqzeroone({\cal C})$ if and only if $S$ is equivalent
to $\bigwedge_{1 \leq i,j \leq n} A(x_i, x_j)$.

It remains to consider the case where every constraint in ${\cal C}$
is complementive. Let $V_0$ be the set of constraint applications of
${\cal C}$ with constant $0$ from Lemma~\ref{l:CHkeylemma}. Let $V_f$
be the set of constraint applications of ${\cal C}$ that results when
we replace each occurrence of 0 in $V_0$ by $f$, where $f$ is a new
variable. Note that the following holds.  There are two cases to
consider, depending on the form of $V_0$.
\begin{description}
\item[Case 1:] $V_0(x, y)$ is equivalent to $(x \rightarrow y)$. In
this case, consider $V_f(f,x,y)$.  Since $V_f(0, x, y)$ is equivalent
to $x \rightarrow y$, and every constraint in $S$ is complementive, it
follows that $V_f(f, x, y)$ is equivalent to $(\overline{f} \wedge (x
\rightarrow y)) \vee ({f} \wedge (y \rightarrow x))$.  Thus,
$\bigwedge_{i = 1}^n x_i \vee \bigwedge_{i = 1}^n \overline{x_i}$ is
equivalent to $\bigwedge_{1 \leq i,j \leq n} V_f(f, x_i ,x_j)$, and it
follows that $S \not \in \SATneqzeroone({\cal C})$ if and only if $S$
is equivalent to $\bigwedge_{1 \leq i,j \leq n} V_f(f, x_i ,x_j)$.
\item[Case 2:] $V_0(x, y, z)$ is equivalent to $(\overline{x} \wedge
\overline{y} \wedge \overline{z}) \vee (x \wedge \overline{y} \wedge
z) \vee (\overline{x} \wedge y \wedge z)$.  Since all constraints in
$V_0$ are complementive, $V_f(f, x, y, z)$ behaves as follows: $V_f(0,
0, 0, 0) = V_f(0, 1, 0, 1) = V_f(0, 0, 1, 1) = V_f(1, 1, 1, 1) =
V_f(1, 0, 1, 0) = V_f(1, 1, 0, 0) = 1$, and $V_f$ is 0 for all other
assignments. Note that $V_f(f, f, x_i, x_j)$ is equivalent to $(x_i
\leftrightarrow x_j)$, and thus that $\bigwedge_{i = 1}^n x_i \vee
\bigwedge_{i = 1}^n \overline{x_i}$ is equivalent to $\bigwedge_{1
\leq i,j \leq n} V_f(f, f, x_i ,x_j)$.  It follows that $S \not \in
\SATneqzeroone({\cal C})$ if and only if $S$ is equivalent to
$\bigwedge_{1 \leq i,j \leq n} V_f(f, f, x_i ,x_j)$.
\qed
\end{description}

\section{The Isomorphism Problem for Constraint Satisfaction}
\label{s:iso}

In this section, we study a more general problem:
The question of whether a set of constraint applications
can be made equivalent to a second set of constraint applications
using a suitable renaming of its variables. We need some definitions.

\begin{definition}\rm
\begin{enumerate}
\item Let $X = \set{\enu{x}{1}{n}}$ be a set of variables. By
$\pi\colon\set{\enu{x}{1}{n}} \rightarrow \set{\enu{x}{1}{n}}$ we
denote a \emph{permutation} of $X$.
\item Let $S$ be a set of constraint applications over variables $X$
and $\pi$ a permutation of $X$.  By \emph{$\pi(S)$}
we denote the set of constraint applications that results when we
replace simultaneously all variables $x_i$ of $S$ by $\pi(x_i)$. 
\item Let $S$ be a set of constraint applications over variables $X$.
The number of satisfying assignments of $S$ is
$\cnto{S} \eqd ||\set{\,I 
\condition I \text{ is an assignment to all variables in }\allowbreak
X \text{ that satisfies every }\allowbreak\text{constraint}\allowbreak \text{application
in $S$\,}}||$.
\end{enumerate}
\end{definition}

The isomorphism problem now is formally defined as follows:

\begin{definition}\rm
\begin{enumerate}
\item $\ISO({\cal C})$ is the problem of, given two sets $S$ and $U$
of constraint applications of ${\cal C}$ over variables $X$,
to decide whether $S$ and $U$ are isomorphic, i.e.,
there exists a permutation $\pi$ of $X$ such that $\pi(S)$ is
equivalent to $U$.
\item $\ISO_c({\cal C})$ is the problem of, given two sets $S$ and $U$
of constraint applications of ${\cal C}$ with constants over
variables $X$, to decide
whether $S$ and $U$ are isomorphic.
\end{enumerate}
\end{definition}

We remark that for $S$ and $U$ to be isomorphic, we require that
formally they are defined over the same set of variables. Of course,
this does not mean that all these variables actually have to occur
textually in both formulas.
 
As in the case for equivalence, isomorphism is in some sense as least as hard
as non-satisfiability, since $S$ is not satisfiable if and only
if $S$ is isomorphic to $0$.  Thus, we immediately obtain that
if ${\cal C}$ is not Schaefer, then $\ISO_c({\cal C})$ is coNP-hard.
Unlike the equivalence case however, we do {\em not} have a trivial coNP
upper bound for isomorphism problems.
In fact, there is some evidence~\cite{AgTh00} that the
isomorphism problem for Boolean formulas is not in coNP.
Note that determining whether two formulas or two
sets of constraint applications are isomorphic is trivially in
$\Sigma_2^p$. However, the isomorphism problem for formulas
is not $\Sigma_2^p$-complete unless the polynomial
hierarchy collapses~\cite{AgTh00}. In the sequel (Theorem~\ref{t:iso-upper})
we will prove a stronger result for the isomorphism problem for Boolean
constraints: We will prove a $\parallelnp$ upper bound for these
problems, where $\parallelnp$ is the class of problems that can be solved
via parallel access to NP.  This class
has many different characterizations, see, for example,
Hemaspaandra~\cite{Hem89},
Papadimitriou and Zachos~\cite{PaZa83},
Wagner~\cite{Wag90}. 

For equivalence, we obtained a polynomial-time upper bound for sets
of constraints that are Schaefer.
In contrast, as we will show in the sequel,
it is easy to see that, for example, isomorphism
for positive 2CNF formulas (i.e., 
isomorphism between two sets of constraint applications
of $\{(0,1), (1,0), (1,1)\}$)
is polynomial-time many-one equivalent to the graph isomorphism problem (\GI).

The main result of this section is the following theorem.

\begin{theorem}\label{isodichth}
Let ${\cal C}$ be a set of constraints.  If ${\cal C}$ is Schaefer,
then $\ISO({\cal C})$ and $\ISO_c({\cal C})$ are polynomial-time
many-one reducible to \GI, otherwise, $\ISO({\cal C})$ and
$\ISO_c({\cal C})$ are {\rm coNP}-hard.
\end{theorem}

Note that if $\cal C$ is Schaefer, then the isomorphism problem
$\ISO({\cal C})$ cannot be coNP-hard, unless the polynomial-time
hierarchy collapses. (This follows since, by our theorem, if
$\ISO({\cal C})$ is coNP-hard then GI is coNP-hard
and, since ${\rm GI}\in {\rm NP}$, coNP would be a subset of NP and thus
NP=coNP which implies the mentioned collapse.) Under the
assumption that the polynomial-time hierarchy does not collapse,
Theorem~\ref{isodichth} thus distinguishes an easy case (reducible to
GI) and a hard case. In this sense, Theorem~\ref{isodichth} is again a
dichotomy theorem.

We will first have a look at the lower bound part of Theorem~\ref{isodichth}.
For that we need the following property:

\begin{lemma}
\label{IsoCnt}
Let $S$ and $U$ be sets of constraint applications of ${\cal{C}}$
with constants.
If $S$ is isomorphic to $U$ then $\cnto{U} = \cnto{S}$.
\end{lemma}

\proof
First note that every permutation of the variables of $S$ induces a
permutation of the rows of the truth-table of $S$. 
Now let $\pi$ be a permutation such that $\pi(S) \equiv U$. Then
$\cnto{S} = \cnto{\pi(S)}$ and $\cnto{\pi(S)} = \cnto{U}$.
\qed

\begin{theorem}\label{isohard}
If ${\cal C}$ is not Schaefer, then $\ISO({\cal C})$ is {\rm
coNP}-hard.
\end{theorem}

\proof 
We first note that a claim analogous to Claim~\ref{NotZOVal} also
holds for isomorphism, i.e., if ${\cal C}$ is not Schaefer, not
0-valid, and not 1-valid, then $\ISO({\cal C})$ is coNP-hard. For the
proof, we use the same reduction as in the proof of
Claim~\ref{NotZOVal} and claim that, again, $S \in
\overline{\SAT({\cal C})}$ if and only if $S$ is isomorphic to $\{C_0(y,
\dots, y), C_1(y, \dots, y)\}$.  For the direction from left to right
note that if $(S,\allowbreak \set{C_0(y, \dots, y),\allowbreak C_1(y,
\dots, y)}) \in \EQUIV({\cal C})$ then also $(S,\allowbreak
\set{C_0(y, \dots, y),\allowbreak C_1(y, \dots, y)}) \in \ISO({\cal
C})$ by the identity permutation $\pi = \textrm{id}$. For the other
direction note that $S \not\in \overline{\SAT({\cal C})}$ iff
$\cnto{S} > 0$. Now suppose $(S, \set{C_0(y, \dots, y), C_1(y, \dots,
y)}) \in \ISO({\cal C})$, then by Lemma~\ref{IsoCnt} $\cnto{S} =
\cnto{\set{C_0(y, \dots, y), C_1(y, \dots, y)}}$, which is clearly a
contradiction since $\cnto{\set{C_0(y, \dots, y), C_1(y, \dots, y)}}=
0$.

Next, we claim, analogously to Claim~\ref{NotZoOVal}, that
\begin{enumerate}[(1)]
\item if ${\cal C}$ is 1-valid, not Schaefer, and not 0-valid, then
$\ISO({\cal C})$ is {\rm coNP}-hard; and
\item If ${\cal C}$ is 0-valid, not Schaefer, and not 1-valid, then
$\ISO({\cal C})$ is {\rm coNP}-hard.
\end{enumerate}
For the first case, we use the same reduction as in the proof
of Claim~\ref{NotZoOVal}. Note that
if constraint set $S$ is equivalent to $\set{C(x_i, \dots, x_i) \condition
1 \leq i \leq n}$, then $(S, \set{C(x_i, \dots, x_i) \condition 1 \leq
i \leq n}) \in \ISO({\cal C})$ via $\pi = \textrm{id}$. For the other
direction
note that $S \not\in \overline{\SATneqone({\cal C})}$ iff
$\cnto{S} \ge 2$, but 
$\cnto{\set{C(x_i, \dots, x_i) \condition 1 \leq
i \leq n}} = 1$. By Lemma~\ref{IsoCnt} the result follows.
The proof of the second case is similar.

The remaining case is that of a set ${\cal C}$ not Schaefer, but both
0-valid and 1-valid. We use the same reduction as in
Claim~\ref{cl:zeroonevalid}. Clearly if $\pair{S,U} \in \EQUIV({\cal C})$
then also $\pair{S,U} \in \ISO({\cal C})$ via $\pi = \mathrm{id}$. To show
the other direction note that if $S \not\in \SATneqzeroone({\cal C})$
then $\cnto{S} \ge 3$, but $\cnto{\bigwedge_{i = 1}^n x_i \vee
\bigwedge_{i = 1}^n \overline{x_i}} = 2$. Now use Lemma~\ref{IsoCnt}
to show that $S$ is not isomorphic to one of the formulas constructed
in the cases examined in Claim~\ref{cl:zeroonevalid}. This completes
the proof of the theorem.
\qed

\medskip

To complete the proof of Theorem~\ref{isodichth}, it remains to show
that if ${\cal C}$ is Schaefer, then $\ISO({\cal C})$ and
$\ISO_c({\cal C})$ are polynomial-time many-one reducible to \GI.  We
will reduce $\ISO_c({\cal C})$ to graph isomorphism for vertex-colored
graphs, a \GI\ variation that is polynomial-time many-one equivalent
to \GI.

\begin{definition}
\VCGI\ is the problem of, given two vertex-colored graphs $G_i = (V_i,
E_i, \chi_i)$, $i \in \{1, 2\}$, $\chi_i: V \rightarrow {\bf N}$, to
determine whether there exists an isomorphism between $G_1$ and $G_2$
that preserves colors, i.e.,\ whether there exists a bijection
$\pi\colon V_1 \rightarrow V_2$ such that $\{v, w\} \in E_1$ iff
$\{\pi(v), \pi(w)\} \in E_2$ and $\chi(v) = \chi(\pi(v))$.
\end{definition}

\begin{proposition}[\cite{F76,BC79}]
\label{vcgith}
\VCGI\ is polynomial-time many-one equivalent to \GI.
\end{proposition}

By Proposition~\ref{vcgith}, to complete the proof of
Theorem~\ref{isodichth}, it suffices to show the following.

\begin{theorem}\label{isoeasy}\label{isotogith}
Let ${\cal C}$ be a set of constraints.  If ${\cal C}$ is Schaefer,
then $\ISO_c({\cal C}) \manyone \VCGI$.
\end{theorem}


\proof Suppose ${\cal C}$ is Schaefer, and let $S$ and $U$ be sets of
constraint applications of ${\cal C}$ with constants over
variables $X$. We will first
bring $S$ and $U$ into normal form.

Let $\widehat{S}$ be the set of all constraint applications $A$ of ${\cal
C}$ with constants such that all of $A$'s variables occur in $X$ and
such that $S \rightarrow A$. Similarly, let $\widehat{U}$ be the set
of all constraint applications $B$ of ${\cal C}$ with constants such
that all of $B$'s variables occur in $X$ and such that $U \rightarrow
B$.  It is clear that $S \equiv \widehat{S}$, since $S \subseteq
\widehat{S}$ and $S \rightarrow \widehat{S}$. Likewise, $U \equiv
\widehat{U}$. Note that $\widehat{S}$ and $\widehat{U}$ are
polynomial-time computable (in $|\pair{S,U}|$), since
\begin{enumerate}
\item there exist at most $|{|\cal C} ||(||X|| + 2)^k$ constraint
applications $A$ of ${\cal C}$ with constants such that all of $A$'s
variables occur in $X$, where $k$ is the maximum arity of constraints
in ${\cal C}$; and
\item since ${\cal C}$ is Schaefer, determining whether $S \rightarrow
A$ or $U \rightarrow A$ takes polynomial time, using the same argument
as in the proof of Theorem~\ref{equivttsatth}.
\end{enumerate}
Note that we have indeed brought $S$ and $U$ into normal form, since
if $S \equiv U$ then $\widehat{S} = \widehat{U}$,
so for any permutation $\pi$ of $X$, if $\pi(S) \equiv U$,
then $\pi(\widehat{S}) =
\widehat{U}$. We remark that this approach of first bringing the sets
of constraint applications into normal form is also followed in the
${\rm coIP}[1]^{\rm NP}$ upper bound proof for isomorphism between
Boolean formulas~\cite{AgTh00}.

It remains to show that we can in polynomial time encode $\widehat{S}$
and $\widehat{U}$ as vertex-colored graphs $G(\widehat{S})$ and
$G(\widehat{U})$ such that there exists a permutation $\pi$ of $X$
with $\pi(\widehat{S}) = \widehat{U}$ if and only if
$\pair{G(\widehat{S}), G(\widehat{U})} \in \VCGI$.

Let ${\cal C} = \{C_1, \ldots, C_m\}$,
let $P = \{C_{i_1}(x_{11}, x_{12},\ldots, x_{1k_1}),\allowbreak
C_{i_2}(x_{21}, x_{22}, \ldots ,  x_{2k_2}),\allowbreak \ldots,\allowbreak
C_{i_\ell}(x_{\ell 1}, x_{\ell 2}, \ldots, x_{\ell k_\ell})\}$ be
a set of constraint applications of ${\cal C}$ with constants
over variables $X$
such that $i_1 \leq i_2 \leq i_3 \leq \cdots \leq i_{\ell}$.
Define $G(P) = (V, E, \chi)$ as the following vertex-colored graph:
\begin{itemize}
\item $V = \{0,1\} \cup \{\,x \condition x \in X\,\} \cup
\{\,a_{ij} \condition 1 \leq i \leq \ell, 1 \leq j \leq k_i\,\}
\cup \{\,A_i \condition 1 \leq i \leq \ell\,\}$. That is,
the set of vertices corresponds to the Boolean constants,
the variables in $X$, the arguments of the constraint applications in $P$, and 
the constraint applications in $P$.
\item $E = \{\,\{x, a_{ij}\} \condition x = x_{ij}\,\} \cup
\{\,\{a_{ij}, A_i\} \condition 1 \leq i \leq \ell, 1 \leq j \leq k_i\,\}.$
\item The vertex coloring $\chi$ will distinguish the different
categories.  Of course, we want to allow any permutation
of the variables, so we will give all elements of $X$ the same color.
In addition, we also need to allow a permutation of constraint applications
of the same constraint.
\begin{itemize}
\item $\chi(0) = 0$, $\chi(1) = 1$,
\item $\chi(x) = 2$ for all $x \in X$,
\item $\chi(A_r) = 2 + j$ if $i_r = j$, and
\item $\chi(a_{ij}) = 2 + m + j$.  (This will ensure that we do
not permute the order of the arguments.) 
\end{itemize}
\end{itemize}
If there is a permutation $\pi$ of $X$ such that 
$\pi(\widehat{S}) = \widehat{U}$,
it is straightforward to see that $\pair{G(\widehat{S}),
G(\widehat{U})} \in \VCGI$.
On the other hand, if $\pair{G(\widehat{S}),
G(\widehat{U})} \in \VCGI$ via a permutation $\pi$ of the vertices of
$G(\widehat{S})$, then note that vertices corresponding to constraint applications
can only be permuted together with those vertices corresponding to the
arguments of that constraint application. In addition, because of the
coloring, the order of arguments is preserved.
Thus, if $\pi(A_i)=A_j$ then necessarily $\pi(a_{ir})=a_{jr}$,
for all $1\leq r \leq k_i$ and $A_i$ and, because coloring is preserved,
$A_i$ and $A_j$ are instances of the same constraint. 
This part of the permutation corresponds to a permutation of the
constraint applications in the set $\widehat{S}$. The remaining
part of the permutation in $G(\widehat{S})$ is one that solely
permutes vertices corresponding to variables in $\widehat{S}$, so
$\pi(\widehat{S}) = \widehat{U}$.
\qed
 

Note that the construction used in proof of the previous theorem
can be used to provide a general upper bound on $\ISO_c({\cal C})$:
Given sets $S$ and $U$ of constraint applications of ${\cal C}$ with constants,
first bring $S$ and $U$ into the normal form ($\widehat{S}$ and $\widehat{U}$)
described in the proof of the previous theorem
(this can be done in polynomial time
with parallel access to an NP oracle), and then determine
if there exists a permutation $\pi$ such that $\pi(\widehat{S})$ =
$\widehat{U}$ (this takes one query to an NP oracle).  The whole 
algorithm takes polynomial time with two rounds of parallel queries to NP,
which is equal to $\parallelnp$ (Buss and Hay~\cite{BusHay91}).
Thus, we have the following
upper bound on the isomorphism problem for contraint satisfaction:

\begin{theorem}
\label{t:iso-upper}
Let ${\cal C}$ be a finite set of contraints.
$\ISO({\cal C})$ and $\ISO_c({\cal C})$ are in
$\parallelnp$.
\end{theorem}

Finally, we show that for some simple instances of Horn, bijunctive, and
affine constraints, the isomorphism problem is in fact polynomial-time
many-one equivalent to the graph isomorphism problem. Proofs of these
results will be given in the appendix.

\begin{theorem}
\label{giequivth}
\GI\ is polynomial-time many-one equivalent 
to $\ISO(\{\{(0,1),\allowbreak(1,0),\allowbreak(1, 1)\}\})$ and to
$\ISO_c(\{\{(0,1),\allowbreak(1,0),\allowbreak(1, 1)\}\})$.
\end{theorem}

\proof
It suffices to show that $\GI
\manyone \ISO(\{\{(0,1), (1,0), (1, 1)\}\})$, since, by
Theorem~\ref{isotogith}, $\ISO_c(\{\{(0,1), (1,0), (1, 1)\}\}) \manyone
\GI$.

Let $G = \pair{V,E}$ be a graph and let
$V = \{1, 2, \ldots , n\}$. We encode $G$ in the 
obvious way as a set of constraint applications:
$S(G) = \{x_i \vee x_j \condition 
\{i,j\} \in E\}$. It is immediate that if $G$ and $H$ are two graphs
with vertex set $\{1, 2, \ldots, n\}$, then
$G$ is isomorphic to $H$ if and only if
$S(G)$ is isomorphic to $S(H)$.
\qed

Note that the constraint $\{(0,1), (1,0), (1, 1)\}$ is the binary
constraint $x \vee y$, denoted by ${\rm OR}_0$ in
\cite{CrKhSu00}. Theorem~\ref{giequivth} can alternatively be
formulated as: \GI\ is polynomial-time many-one equivalent the
isomorphism problem for positive 2CNF formulas (with or without
constants). Also, from \cite{To00}, we conclude that this isomorphism
problem thus is hard for NL, PL, and DET.

\begin{theorem}
\label{t:gi-affine}
\GI\ is polynomial-time many-one equivalent to
$\ISO($$\{\{(1,0,0),\allowbreak
(0,1,0),\allowbreak(0,0,1),\allowbreak(1,1,1)\}\})$ and to
$\ISO_c(\{\{(1,0,0),\allowbreak(0,1,0),\allowbreak(0,0,1),\allowbreak
(1,1,1)\}\})$.
\end{theorem}

\proof
It suffices to show that $\GI
\manyone
\ISO(\{\{(1,0,0), (0,1,0), (0,0,1), (1,1,1)\}\})$, since, by
Theorem~\ref{isotogith},
$\ISO_c(\{\{(1,0,0), (0,1,0), (0,0,1), (1,1,1)\}\}) \manyone \GI$.

Let $G = \pair{V,E}$ be a graph, let
$V = \{1, 2, \ldots , n\}$, and enumerate the edges
as $E = \{e_1, e_2, \ldots, e_m\}$.
We encode $G$ as a set of ${\rm XOR}_3$ constraint applications in which
propositional variable $x_i$ will correspond to vertex $i$
and propositional variable $y_i$ will correspond to edge $e_i$.
We encode $G$ as $S(G) = S_1(G) \cup S_2(G) \cup S_3(G)$ where
\begin{itemize}
\item $S_1(G) = \{ x_i \xor x_j \xor y_k \ | \ e_k = \{i,j\}\}$
($S_1(G)$ encodes the graph),
\item $S_2(G) = \{ x_i \xor z_i \xor z'_i \ | \ i \in V\}$
($S_2(G)$ will be used to distinguish $x$ variables
from $y$ variables),
and
\item $S_3(G) = \{ y_i \xor y_j \xor y_k \ | \
e_i, e_j, \mbox{and } e_k \mbox{ form a triangle in } G\}$.
Note that for every $A \in S_3(G)$, $S_1(G) \rightarrow A$.
We add these constraint applications to $S(G)$ to ensure that
$S(G)$ is a maximum set of ${\rm XOR}_3$ formulas.
\end{itemize}

We will show later that if $G$ and $H$ are two graphs
with vertex set $\{1,2, \ldots, n\}$ 
without isolated vertices, then
$G$ is isomorphic to $H$ if and only if
$S(G)$ is isomorphic to $S(H)$. 

The proof of the theorem relies on the following lemma, which 
shows that $S(G)$ is a maximum set of ${\rm XOR}_3$ formulas.  This is an
important property, since checking whether two maximum sets
of functions are equivalent basically amounts to checking whether the
sets are equal, as will be explained in detail after the
proof of the lemma.

\begin{lemma}
\label{l:maxxor}
Let $G = \pair{V,E}$ be a graph such that
$V = \{1,2, \ldots, n\}$ and $E = \{e_1,e_2, \ldots, e_m\}$.
Then for every triple of distinct propositional variables $a,b,c$
in $S(G)$, the following holds: If $S(G) \rightarrow a \xor b \xor c$, then
$a \xor b \xor c \in S(G)$.  Note: we view $a \xor b \xor c$ as a 
{\em function}, and thus, for example,
$a \xor b \xor c = c \xor a \xor b$.
\end{lemma}

\proof
Suppose that there exists a triple of distinct
propositional variables $a$, $b$, and $c$ in $S(G)$ such that
$S(G) \rightarrow a \xor b \xor c$ and
$a \xor b \xor c \not \in S(G)$.
Let $X = \{x_i \ | \ i \in V\}$, $Y = \{y_i \ | \ e_i \in E\}$, and 
$Z = \{z_i, z'_i \ | \ i \in V\}$.
Without loss of generality,
assume that $a \leq b \leq c$, where $\leq$ is the following order 
on $X \cup Y \cup Z$:
\[x_1 < \cdots < x_n < y_1 < \cdots < y_m < z_1 < \cdots < z_n < z'_1 < \cdots
< z'_n.\] 

The proof consists of a careful analysis of different sub-cases.
We will show that in each case, there exists an assignment 
on $X \cup Y \cup Z$ such that
that satisfies $S(G)$  but not $(a \xor b \xor c)$,
which contradicts the assumption that
$S(G) \rightarrow a \xor b \xor c$.

It is important to note that any assignment to $X$ can be extended
to a satisfying assignment of $S(G)$.

\begin{enumerate}
\item If exactly three or exactly one
of the variables in  $\{a, b,c\}$ are in $Z$,
then consider the assignment
that assigns $0$ to every variable in $Z$ and 
$1$ to every variable in $X \cup Y$. Clearly, this assignment
satisfies $S(G)$ but not $(a \xor b \xor c)$.

\item  If exactly two of the variables
in  $\{a, b, c\}$ are in $Z$, then, since $a < b < c$,
$b$ and $c$ are in $Z$. We consider the 
the following two sub-cases, depending on whether $a \in X$ or $a \in Y$.
\begin{enumerate}
\item 
If $a \in X$, then set $b$ and $c$ to $1$ and $a$ to $0$.
Since by assumption $a \xor b \xor c \not \in S(G)$,
it is easy to see that 
this assignment can be extended to an assignment
on $X \cup Z$ that satisfies 
$x_i \xor z_i \xor z'_i$ for all $i \in V$.
This assignment in turn can be extended to an assignment
on $X \cup Y \cup Z$ that also satisfies every constraint application of
the form $x_i \xor x_j \xor y_k$ for $e_k = \{i,j\}$.
So, we now have an assignment that satisfies
$S_1(G)$ and $S_2(G)$. Since
$S_1(G) \rightarrow A$ for every constrain application $A \in S_3(G)$,
it follows that this assignment also satisfies $S(G)$ while
it does not satisfy $(a \xor b \xor c)$.

\item If $a \in Y$,  let $a = y_k$ where $e_k = \{i,j\}$.
If $\{b,c\} = \{z_\ell,z'_\ell\}$ for some $\ell$, then 
we set $x_{\ell}$ to $1$. In all cases, set exactly one
of $\{x_i,x_j\}$ to $1$ (this could be $x_\ell$). Set all
other elements of $X$ to $0$. 
We can extend this to
a satisfying assignment of $S(G)$ that does not satisfy
$(a \xor b \xor c)$.
\end{enumerate}

\item If $a$, $b$, and $c$ are in $X$, then
set $a$, $b$, and $c$ to $0$. It is easy to see that
this assignment can be extended to an assignment
on $X \cup Z$ that satisfies 
$x_i \xor z_i \xor z'_i$ for all $i \in V$.
This assignment in turn can be extended to an assignment
on $X \cup Y \cup Z$ that also satisfies 
$x_i \xor x_j \xor y_k$ for $e_k = \{i,j\}$.
So, we now have an assignment that satisfies
$S(G)$  but does not satisfy $(a \xor b \xor c)$.

\item If $c \in Y$ and $a$ and $b$ are in $X$,
suppose that $c = y_k$ and
let $e_k = \{i,j\}$.  By the assumption that $a \xor b \xor c$
is not in $S(G)$, at least one of $a$ and $b$ is not
in $\{x_i,x_j\}$.

Without loss of generality, let $a \not \in
\{x_i,x_j\}$. Set $a$ to 0 and set $X \setminus \{a\}$ to $1$.
This assignment can be extended to a satisfying assignment for
$S(G)$. Note that such an assignment will set $y_k$ to $1$.
It follows that this assignment does not satisfy $a \xor b \xor c$. 

\item If $a \in X$ and $b$ and $c$ are in $Y$, then
set $a$ to $0$ and $b$ and $c$ to $1$. It is easy to see that
this can be extended to a satisfying assignment for $S(G)$.

\item If  $a$, $b$, and $c$ are in $Y$, let $a = y_{k_1}, b = y_{k_2},
c = y_{k_3}$ such that $e_{k_\ell} = \{i_{\ell},j_{\ell}\}$
for $\ell \in \{1,2,3\}$.
First suppose that for every $\ell \in \{1, 2, 3\}$, for every
$x \in \{x_{i_{\ell}},x_{j_{\ell}}\}$, there exists an
$\ell' \in \{1,2,3\}$ with  $\ell' \neq \ell$ and a constraint
application $A$ in $S(G)$ such that $x$ and $y_{k_{\ell'}}$ occur in $A$. 
This implies that every vertex in 
$\{i_1, j_1, i_2, j_2, i_3, j_3\}$ is incident with at least
2 of the edges in $e_{k_1}, e_{k_2}, e_{k_3}$.
Since these three edges are distinct, it
follows that the edges $e_{k_1}, e_{k_2}, e_{k_3}$ form a triangle
in $G$, and thus $y_{k_1} \xor y_{k_2} \xor y_{k_3} \in S(G)$.
This is a contradiction.

So, let $\ell \in \{1,2,3\}$, $x \in \{x_{i_{\ell}},x_{j_{\ell}}\}$
be such that for all $\ell' \in \{1,2,3\}$ with $\ell \neq \ell'$,
$x$ and $y_{k_{\ell'}}$ do not occur in the same constraint application 
in $S(G)$.  Set $x$ to $0$ and set $X \setminus \{x\}$ to $1$. This can be
extended to a satisfying assignment of $S(G)$ and such a 
satisfying assignment must have the property that $y_{k_{\ell}}$ is $0$ and
$y_{k_{\ell'}}$ is $1$ for all $\ell' \in \{1,2,3\}$ such that
$\ell' \neq \ell$.
\qed
\end{enumerate}

How can Lemma~\ref{l:maxxor} help us in the proof of
Theorem~\ref{t:gi-affine}?
Note that if $S$ and $T$ are maximum sets of ${\cal C}$ constraint
applications, 
then $S \equiv T$ if and only if $S = T$.  Here equality should be
seen as equality between sets of functions, i.e., 
$a \xor b \xor c = b \xor c \xor a$ etc.
So $S$ is isomorphic to $T$ if and only if there exists
a permutation $\rho$ of the variables of $S$ such that
$\rho(S) = T$.

We will now prove Theorem~\ref{t:gi-affine}.
Let $G$ and $H$ be two graphs.  Remove the isolated vertices from
$G$ and $H$. If $G$ and $H$ thus modified do not have the same
number of vertices or they do not have the same
number of edges, then $G$ and $H$ are clearly not isomorphic.
If $G$ and $H$ have the same number of vertices
and the same number of edges, then rename
the vertices in such a way that the vertex set of both graphs
is $V = \{1,2,\ldots,n\}$. Let $\{e_1, \ldots, e_m\}$ be an enumeration
of the edges in $G$ and let $\{e'_1, \ldots, e'_m\}$ be an
enumeration of the edges in $H$.

We will show that $G$ is isomorphic to $H$ if and only if
$S(G)$ is isomorphic to $S(H)$. 

The left-to-right direction is trivial, since an isomorphism between
the graphs induces an isomorphism between sets of constraint
applications as follows.
If $\pi: V \rightarrow V$ is an isomorphism from $G$ to $H$, then
$\rho$ is an isomorphism from $S(G)$ to $S(H)$ defined as follows:
\begin{itemize}
\item
$\rho(x_i) = x_{\pi(i)}$,  
$\rho(z_i) = z_{\pi(i)}$,
$\rho(z'_i) = z'_{\pi(i)}$, for $i \in V$.
\item 
For $e_k = \{i,j\}$, 
$\rho(y_k) = y_{\ell}$ where $e'_{\ell} = \{\pi(i),\pi(j)\}$.
\end{itemize}

For the converse, suppose that $\rho$ is an isomorphism
from $S(G)$ to $S(H)$.  By the observation above, $\rho(S(G)) = S(H)$.
Now look at the properties of the different classes of variables.
\begin{enumerate}
\item Elements from $X$ are exactly those variables that occur at least
twice and that also occur in an element of $S(G)$ together with two
variables that occur exactly once. So, $\rho$ will map $X$
onto $X$.
\item Elements of $Z$ are those variables that occur exactly once
and that occur together with an element from $X$ and another
element that occurs exactly one. So $\rho$ will map
$Z$ to $Z$.
\item Everything else is an element of $Y$. So, $\rho$ will map
$Y$ onto $Y$.
\end{enumerate}
For $i \in V$, define $\pi(i) = j$ iff $\rho(x_i) = x_j$.
$\pi$ is 1-1 onto by observation (1) above. It remains to show that
$\{i,j\} \in E$ iff $\{\pi(i),\pi(j)\} \in E'$.
Let $e_k = \{i,j\}$. Then $x_i \xor x_j \xor e_k \in S(G)$.
Thus, $\rho(x_i) \xor \rho(x_j) \xor \rho(y_k) \in S(H)$.
That is, $x_{\pi(i)} \xor x_{\pi(j)} \xor \rho(y_k) \in S(H)$.
But that implies that $\rho(y_k) = y_{\ell}$ where 
$e'_{\ell} = \{\pi(i),\pi(j)\}$. This implies that
$\{\pi(i),\pi(j)\} \in E'$. For the converse, suppose that
$\{\pi(i),\pi(j)\} \in E'$. Then
$x_{\pi(i)} \xor x_{\pi(j)} \xor y_{\ell} \in S(H)$ for
$e_{\ell} = \{\pi(i),\pi(j)\}$.  It follows
that $x_i \xor x_j \xor \rho^{-1}(y_{\ell}) \in S(G)$. By the form
of $S(G)$, it follows that $\{i,j\} \in E$.
\qed

Note that the constraint $\{(1,0,0), (0,1,0), (0,0,1), (1,1,1)\})$ is
the constraint $x \xor y \xor z$, denoted by ${\rm XOR}_3$ in
\cite{CrKhSu00}.

Note that the previous proof shows 
$\ISO({\rm NXOR}_3)$ and $\ISO_c({\rm NXOR}_3)$ are many-one
equivalent to GI (just negate all variables).
And we can replace the 3 by any $k \geq 3$
(just use duplicate copies of variables).

{From} Theorems~\ref{giequivth} and \ref{t:gi-affine}, we conclude
that, if we could show that isomorphism for bijunctive, anti-Horn
(and, by symmetry, Horn) or affine CSPs is in P, then the graph
isomorphism problem is in P, settling a long standing open question.

\medskip

\noindent{\bf Acknowledgements:}
We thank Lane Hemaspaandra for helpful conversations and suggestions.

\goodbreak
{\small\bibliography{complexity}}

\newpage

\end{document}